\documentclass[12pt]{spieman}  
\usepackage{amsmath,amsfonts,amssymb}
\usepackage{graphicx}
\usepackage{setspace}
\usepackage{tocloft}
\usepackage{doi}
\usepackage{url}
\title{Visualizing quantum mechanics in an interactive simulation -- Virtual Lab by Quantum Flytrap}

\author[a]{Piotr Migdał}
\author[a]{Klementyna Jankiewicz}
\author[a]{Paweł Grabarz}
\author[b]{Chiara Decaroli}
\author[a]{Philippe Cochin}
\affil[a]{Quantum Flytrap, ul. Ceglowska 29, Warsaw, Poland, 01-809}
\affil[b]{National Quantum Computing Center, Didcot, Oxfordshire, UK, OX11 0GD}

\cftpagenumbersoff{figure}
\cftpagenumbersoff{table} 

\newcommand{\ket}[1]{\left| #1 \right\rangle}
\newcommand{\bra}[1]{\left\langle #1 \right|}
\newcommand{\braket}[2]{\left\langle #1 | #2 \right\rangle}
\newcommand{\projection}[1]{\left| #1 \right\rangle \left\langle #1 \right|}
\newcommand{\bio}[2]{} 

\begin{document} 
\maketitle

\begin{abstract}
\emph{Virtual Lab by Quantum Flytrap} explores novel ways to represent quantum phenomena in an interactive and intuitive way. It is a no-code online laboratory with a real-time simulation of an optical table, supporting up to three entangled photons. Users can place typical optical elements (such as beam splitters, polarizers, Faraday rotators, and detectors) with a drag-and-drop graphical interface. \emph{Virtual Lab} operates in two modes. The sandbox mode allows users to compose arbitrary setups. \emph{Quantum Game} serves as an introduction to \emph{Virtual Lab} features, approachable for users with no prior exposure to quantum mechanics.

We introduce novel ways of visualizing entangled quantum states and displaying entanglement measures, including interactive visualizations of the ket notation and a heatmap-like visualization of quantum operators. These quantum visualizations can be applied to any discrete quantum system, including quantum circuits with qubits and spin chains. These tools are available as open-source \emph{TypeScript} packages -- \emph{Quantum Tensors} and \emph{BraKetVue}.

\emph{Virtual Lab} makes it possible to explore the nature of quantum physics (state evolution, entanglement, and measurement), to simulate quantum computing (e.g. the Deutsch-Jozsa algorithm), to use quantum cryptography (e.g. the Ekert protocol), to explore counterintuitive quantum phenomena (e.g. quantum teleportation \& the Bell inequality violation), and to recreate historical experiments (e.g. the Michelson-Morley interferometer).

\emph{Virtual Lab} is available at: \url{https://lab.quantumflytrap.com}.
\end{abstract}


{\noindent \footnotesize\textbf{*}Piotr Migdał,  \url{piotr@quantumflytrap.com} }

\begin{spacing}{1}   

\section{Introduction}
\label{sect:intro}  

Quantum mechanics is commonly considered an arcane subject, counterintuitive not only to a general audience but even to its own creators. Niels Bohr claimed that \textit{“those who are not shocked when they first come across quantum theory cannot possibly have understood it.”}\cite{heisenberg_physics_1971} Albert Einstein, perplexed by a curious correlation between particles, called it \textit{“a spooky action at a distance“}\cite{einstein_born-einstein_1971} -- a property now called entanglement. Quantum mechanics sparks the imagination in popular culture\cite{kim_physics_2017} and spirituality\cite{kaiser_how_2012}, but is rarely presented in a factual, digestible way. A handful of accessible books are available\cite{scarani_six_2010,susskind_theoretical_2014,matuschak_quantum_2019,wootton_teaching_2021,ekert_online_2022} but still require knowledge of complex vector calculus.

Recently, motivations to learn quantum mechanics went beyond physics education and pure curiosity. Quantum technologies are gaining importance at a rapid pace\cite{bobier_what_2021}, with estimates of more than \$1 trillion value potential by the mid-2030s\cite{hazan_next_2020}. Quantum cryptography is already in proof-of-concept use, while quantum computing will likely add business value in the next few years\cite{waters_goldman_2021}. Consequently, quantum literacy\cite{nita_challenge_2021} is becoming more important in order to keep up with technological possibilities and career opportunities. Therefore, it is crucial to provide practical tools for a wider audience\cite{hughes_assessing_2022}. A similar process occurred for image processing with deep learning, which is now accessible to regular software engineers\cite{migdal_level_2021}.

Part of this need is being bridged by a rapidly growing family of quantum software frameworks\cite{fingerhuth_open_2018,fingerhuth_open-source_2022}, e.g.: \emph{QuTiP}\cite{johansson_qutip_2012}, \emph{Qiskit by IBM}\cite{anis_qiskit_2021}, \emph{Penny Lane}\cite{bergholm_pennylane_2020} \& \emph{Strawberry Fields}\cite{killoran_strawberry_2019} by \emph{Xanadu}, and \emph{Pulser}\cite{silverio_pulser_2022} by \emph{Pasqal}. These packages have various balances of general quantum information, quantum circuits, quantum optimization, plotting capabilities, access to physical implementations of algorithms, and capabilities of running on quantum devices. These frameworks use the \emph{Python} interface, thus allowing usage via the interactive environment of \emph{Jupyter Notebook}. However, recent studies show that programs written in these frameworks are error-prone, with a high percentage of errors in the code being directly related to quantum mechanics\cite{paltenghi_bugs_2021, luo_comprehensive_2022}.

One way of introducing quantum mechanics is through interactive simulations and explorable explanations. For a recent introduction to this approach, see the survey article \textit{“Quantum Games and Interactive Tools for Quantum Technologies Outreach and Education“} by Zeki Seskir et al\cite{seskir_quantum_2022} in which photons are presented as one of the most straightforward introductions to quantum mechanics.
One of the earliest simulations is an interactive visualization of a one-dimensional wavefunction in a potential well\cite{falstad_quantum_2002} from 2002. Newer ones include simulations of quantum computing circuits (\emph{Quirk}\cite{gidney_quirk_2022} and \emph{IBM Quantum Experience}\cite{bozzo-rey_introduction_2018}) and an optical table, \emph{The Virtual Quantum Optics Laboratory (VQOL)} developed at the University of Texas at Austin\cite{la_cour_virtual_2021}.
The \emph{VQOL}’s capabilities and interface focus on realistic laser physics, with the lasers' wavelengths, intensities, and decoherence. While it supports entangled pairs, it does not aim to simulate the delicate properties of a few photons, including interactions (e.g. CNOT gates), nondemolition measurements, or detection-dependent operations. \emph{VQOL} is a complementary tool for simulating and explaining macroscopic systems of light.

Some educational, interactive simulations take the form of games\cite{seskir_quantum_2022}. John Preskill wrote that \textit{“perhaps kids who grow up playing quantum games will acquire a visceral understanding of quantum phenomena that our generation lacks”}\cite{preskill_quantum_2018} and studies confirm that this type of open-ended experience promotes exploration\cite{bonawitz_double-edged_2011}. Children learn Newtonian physics by interacting with their environment\cite{fox_swinging_1997,solis_childrens_2017} rather than starting from differential calculus. We can to provide a similar environment for quantum mechanics. Examples include a citizen science game \emph{Quantum Moves 2}\cite{jensen_crowdsourcing_2021} and a visual quantum circuit puzzle \emph{Quantum Odyssey}\cite{nita_inclusive_2021}.

\emph{Virtual Lab by Quantum Flytrap} is a successor to \emph{Quantum Game with Photons} (2016)\cite{migdal_quantum_2022-1}, an open-source puzzle game with 34 levels and a sandbox. It had over 100k gameplays and was selected as the top pick in gamifying quantum theory in \emph{The Quantum Times}\cite{leifer_gamifying_2017}. We apply a similar drag-and-drop interface with step-by-step dynamics. \emph{Quantum Game with Photons} is restricted to a single photon. Adding more particles required creating a different simulation engine, and creating both numerics and visualizations for entanglement. The graphical user interface of \emph{Virtual Lab} is a complete redesign -- a completely new project, with many ways to interact with and explore quantum experiments. \emph{Quantum Game with Photons} was later developed as \emph{Quantum Game with Photons 2} at the Centre for Quantum Technologies, National University of Singapore\cite{migdal_quantum_2022}.

This paper is organized as follows. In Section~\ref{sec:overview}, we present an overview of \emph{Virtual Lab}. In Section~\ref{sec:design_ui}, we show the core elements of the graphical user interface. Most of these components provide novel ways to interact with quantum states and can be readily used in other quantum interfaces. Specifically, we devised a new way to visualize arbitrary entangled states, and another to show entanglement measures. In Section~\ref{sec:simulation}, we describe the physics of simulation, evolution, and measurement of quantum states. We spell out our assumptions and provide equations for the performed operations. In Section~\ref{sec:experiments}, we show a set of optical experiments that can be directly simulated in \emph{Virtual Lab}. We aim to showcase our interface and simulation capabilities, as well as to provide a set of examples for students, educators, teachers, and lecturers. In Section~\ref{sec:usage}, we provide examples of usage and indicators of popularity. In Section~\ref{sec:conclusion}, we conclude the paper and explore possible directions of development.

\section{Overview}
\label{sec:overview}

\begin{figure}
\begin{center}
\includegraphics[width=0.85\textwidth]{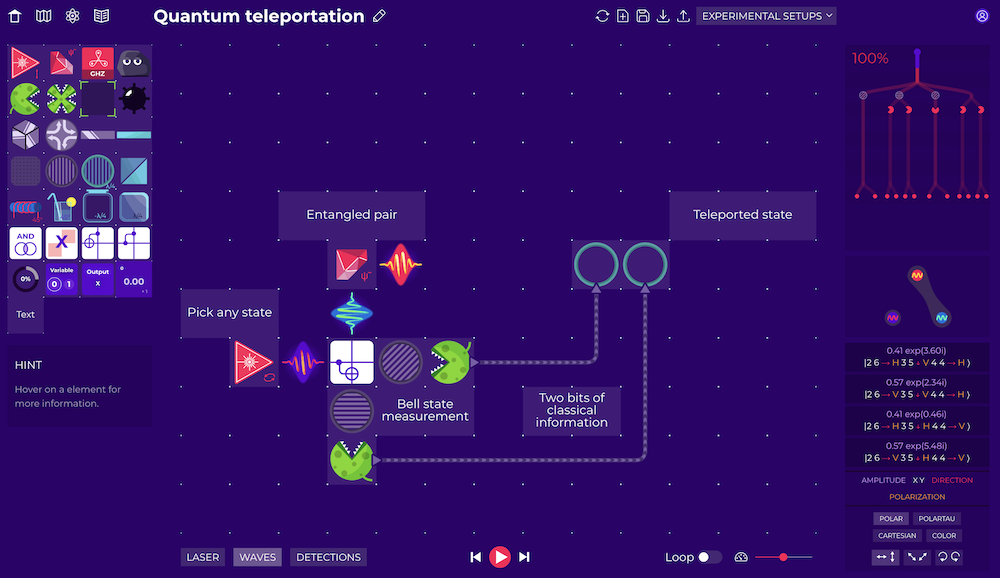}
\end{center}
\caption 
{ \label{fig:virtual_lab}
The main view of \emph{Virtual Lab}: the board (center), the element toolbox (left), experiment controls (bottom), quantum state exploration tools (right), and options for saving and editing experiments (top).} 
\end{figure}

\emph{Virtual Lab by Quantum Flytrap} (available at \url{https://lab.quantumflytrap.com}, see Fig.~\ref{fig:virtual_lab}) is a free-to-use, browser-based laboratory with a real-time simulation of quantum states of photons, the elementary particles of light. \emph{Virtual Lab} is an open-ended experience, which can be used as an educational or research tool, as a puzzle challenge, and as a means for creative expression. For beginners, we aim to lower barriers to entry and serve as their first step to quantum literacy. For educators, \emph{Virtual Lab} is an interactive and intuitive teaching aid. For physicists and quantum software engineers, it offers a way to efficiently investigate advanced quantum-information phenomena and prototype experiments.

\emph{Virtual Lab} is aligned with \emph{Quantum Flytrap}’s goals to bridge the gap between quantum computing and end-users by providing no-code quantum programming interfaces. We have designed \emph{Virtual Lab} to be a highly composable environment (like \emph{LEGO bricks} or \emph{Minecraft}), yet powerful enough to simulate major wave optics and quantum information phenomena. \textit{“Leave computing to computers“} served as our motto.

\emph{Virtual Lab} is based on a numerics engine developed by \emph{Quantum Flytrap}. Unlike prevalent qubit-only abstraction, we focus on concrete hardware implementations -- our engine can be used to simulate all the leading physical realizations including optically-trapped neutral atoms and ions, superconductors, and photons. Since we are now in the Noisy Intermediate Scale Quantum (NISQ) era\cite{preskill_quantum_2018} of early-stage quantum devices, it is crucial to understand the capabilities of concrete physical implementations.

Classical simulations of quantum systems get exponentially more difficult with the number of particles. In our case, a single photon requires 1000 complex numbers (all combinations of its grid position, polarization, and direction). A three-photon simulation requires a billion dimensions. Consequently, \emph{Quantum Flytrap} needed to develop a custom high-performance sparse array simulation in \emph{Rust}, a low-level programming language, as none of the off-the-shelf solutions were suitable. The web interface uses a number of \emph{JavaScript} technologies, including its dialect \emph{TypeScript} and the web framework \emph{Vue.js}. 

\section{Design \& User Interface}
\label{sec:design_ui}

We created \emph{Virtual Lab} as a web-based app to make it accessible to a broad audience. \emph{Virtual Lab}’s install-free nature allows users to instantly get started and facilitates sharing user-created experiments. It can be used on desktop computers and mobile devices with modern browsers, regardless of the operating system.

\begin{figure}
\begin{center}
\includegraphics[width=0.6\textwidth]{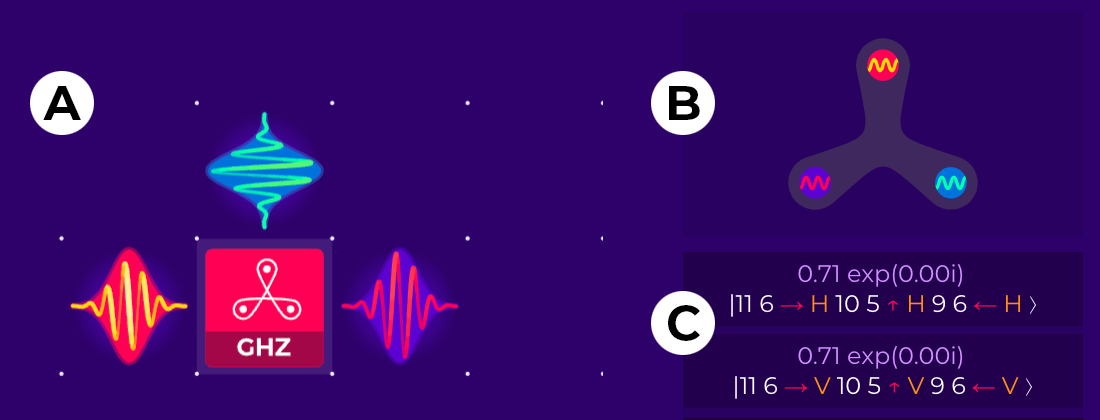}
\end{center}
\caption 
{ \label{fig:ghz}
An experiment in wave mode, which allows the user to explore every step of the simulation. A) Amplitudes visualized as blinking wavepackets, see Sec.~\ref{ss:amp_viz} and ~\ref{ss:amp_entanglement}. B) An entanglement graph, see Sec.~\ref{ss:entanglement_graph}. C) Ket state, see Sec.~\ref{ss:ket}.
} 
\end{figure}

The board is the main component of \emph{Virtual Lab}. It is a grid-based canvas where users can place optical elements and watch the simulation. Elements can be placed with a drag-and-drop interface and rotated with a click. Users can run experiments in three modes,  all simulating precisely the same physics.
The high-intensity mode shows the whole photon path and displays absorption probabilities in real-time, as demonstrated for classical interferometers in Fig.~\ref{fig:experiment_mach_zehnder}.
The wave mode displays amplitudes of photons and offers a way to explore quantum states. This mode can be accessed as a continuous simulation or as a way to examine quantum states at selected time steps, see Fig.~\ref{fig:ghz}.
The detection mode shows a realistic experimental scenario in which we cannot “view” photon amplitudes, but we only receive a discrete signal when a photon has arrived at a detector -- as in quantum mechanics such measurement would affect the state. We use this mode to demonstrate the Bell inequality violation, see ~\ref{fig:experiment_bell_inequality}.

\emph{Virtual Lab} offers a way to save and share experimental setups. Saving experiments requires creating an account. Each experimental setup can be given a name, set to public or unlisted, and shared with a link. Viewing does not require setting up an account -- this choice was made so as to facilitate easier access and distribution. For a list of publicly shared experiments, see:\\
\url{https://lab.quantumflytrap.com/u/}.

\subsection{Optical elements}

\begin{figure}
\begin{center}
\includegraphics[width=0.85\textwidth]{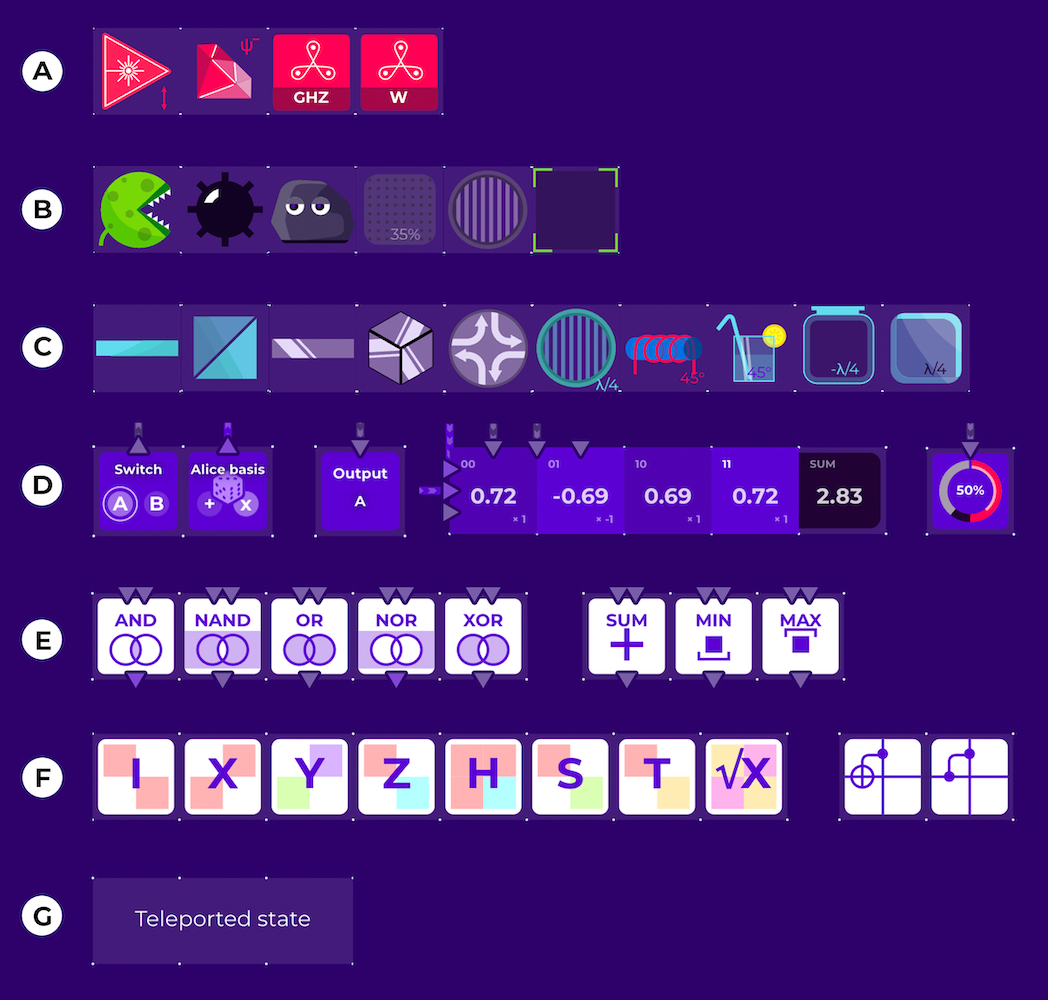}
\end{center}
\caption 
{ \label{fig:elements}
Elements available in \emph{Virtual Lab} grouped by type and described from left-to-right.
A) Photon generators: a single-photon source, a Bell-pair generator, the three-photon GHZ state, and the three-photon W state.
B) Measurements: a detector, a photo-sensitive bomb, a rock, a neutral-density filter, a linear polarizer, a nondemolition detector.
C) Non-absorptive optical elements: a non-polarizing beamsplitter, a polarizing beamsplitter, a mirror, a corner cube, an optical circulator, a waveplate, a Faraday rotator, a sugar solution, a vacuum jar, a glass slab.
D) Input and output: a switch, a randomized switch, an output variable, a correlator, a goal.
E) Classical logic gates: AND, NAND, OR, NOR, XOR; arithmetic operations: SUM, MIN, MAX.
F) Quantum gates\cite{nielsen_quantum_2010,zendejas-morales_quantum_2021} for a single qubit: identity, Pauli-X, Pauli-Y, Pauli-Z, Hadamard, phase shifts, and a square root of NOT. For two qubits: controlled-NOT (CNOT) and controlled-Z.
G) A text comment field.
} 
\end{figure}

\begin{figure}
\begin{center}
\includegraphics[width=0.4\textwidth]{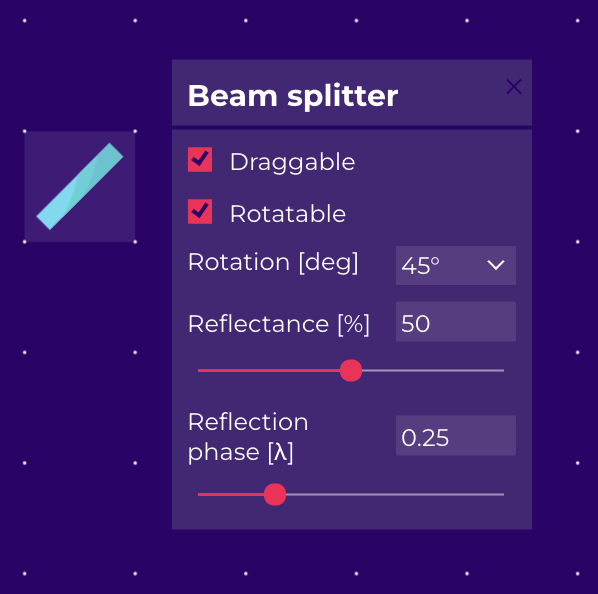}
\end{center}
\caption 
{ \label{fig:element_parameters}
Configurable parameters for a non-polarizing beamsplitter: rotation, reflectance, reflection phase from one of its sides.
} 
\end{figure}

Elements in \emph{Virtual Lab} are based on idealized versions of optical devices that can be used on an optical table in a laboratory. We provide visual icons as symbolic representations of these elements along with logical operators, see Fig.~\ref{fig:elements}. Many elements have additional physical properties, which can be modified in a tooltip that opens after a right-click, see Fig.~\ref{fig:element_parameters}.

The Bell pair generator is based on an abstracted version of the spontaneous parametric down-conversion in a beta barium oxide crystal\cite{kwiat_new_1995}, hence the gem icon. We don't display the incident laser beam and focus only on the generated state.
The detector is represented by a venus flytrap to add playfulness with a recognizable image of the piranha plant from the \emph{Mario} game series. Similarly, we represent a fully-absorbing element as a sentient rock. A photo-sensitive bomb icon takes its inspiration from  another popular game, \emph{Minesweeper}.  We have found that this approach makes \emph{Virtual Lab} a more welcoming introduction to quantum physics and adds recognizability -- to the point that we named the company after this plant.
Each classical logic operation is depicted by a Venn diagram, while a quantum gate -- as a letter on its matrix visualization, see Sec.~\ref{ss:operator}.

In optics, phase delays are typically modified by a slight change of optical path, which is challenging to present with a grid-based system. Therefore, we delay the phase with a glass slab (since it has a higher refraction index than air) and a vacuum chamber (since it has a lower refraction index than air). This representation serves as a further introduction to optical phenomena.
Similarly, in our mirrors, both sides are reflective.
Linear polarizers and phase plates are represented in the board plane, rather than perpendicularly, for visual clarity.
At the same time, we made sure that all elements are expressed with physically-correct operators.

\subsection{Photon amplitude visualization}
\label{ss:amp_viz}

\begin{figure}
\begin{center}
\begin{tabular}{cccc}
\includegraphics[height=0.1\textwidth]{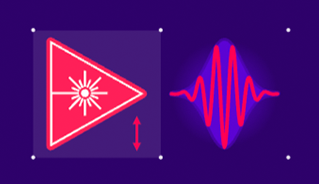} &
\includegraphics[height=0.1\textwidth]{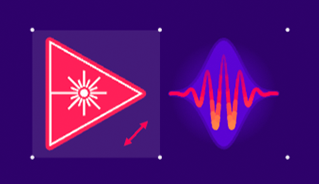} &
\includegraphics[height=0.1\textwidth]{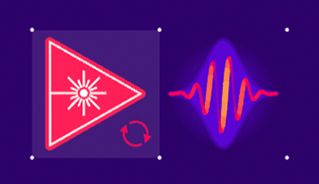} \\
\includegraphics[height=0.1\textwidth]{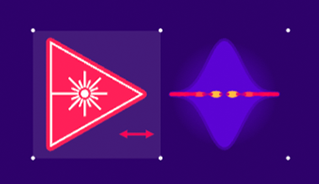} &
\includegraphics[height=0.1\textwidth]{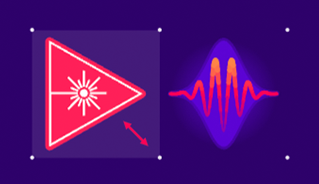} &
\includegraphics[height=0.1\textwidth]{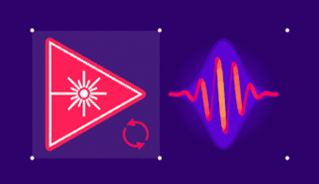} 
\end{tabular}
\end{center}
\caption 
{ \label{fig:amp_viz}
A single-photon amplitude is represented as oscillations within a Gaussian envelope. Take a note that horizontal polarization $\ket{H}$ is parallel to the board, while vertical $\ket{V}$ is perpendicular to the board.} 
\end{figure}

We represent a photon as a wavepacket -- oscillations within a Gaussian envelope, see Fig.~\ref{fig:amp_viz}. We draw an electric field, according to its polarization, in a pseudo-3D way. We show the phase by shifting the oscillation. The amplitude is shown as opacity, rather than more traditionally as the height of the electric field amplitude. We wanted to ensure that small amplitudes are visible and use an intuitive, visual language, which is suitable for a general audience. In our visualizations of vectors and matrices, we use hue to indicate phase. We refrained from the same approach for photons approach, as it would be confusing to use color where it could be confused with the wavelength.

\subsection{Quantum state visualization}
\label{ss:ket}

\begin{figure}
\begin{center}
\begin{tabular}{ccc}
\includegraphics[height=0.3\textwidth]{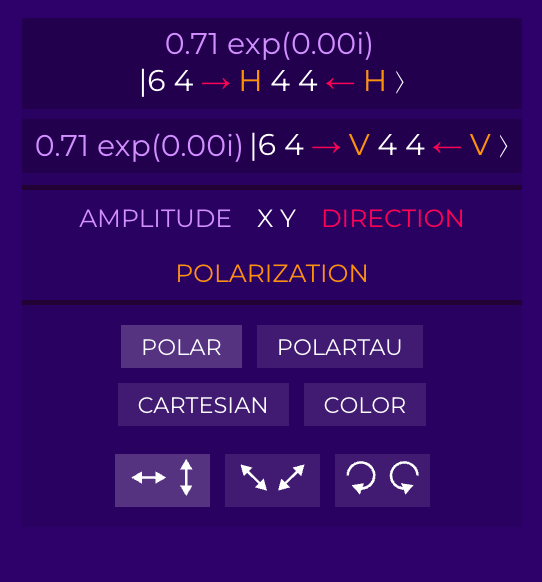} &
\includegraphics[height=0.3\textwidth]{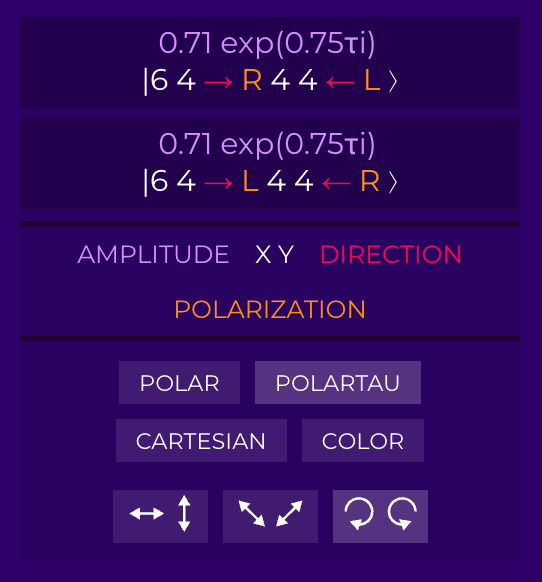} &
\includegraphics[height=0.3\textwidth]{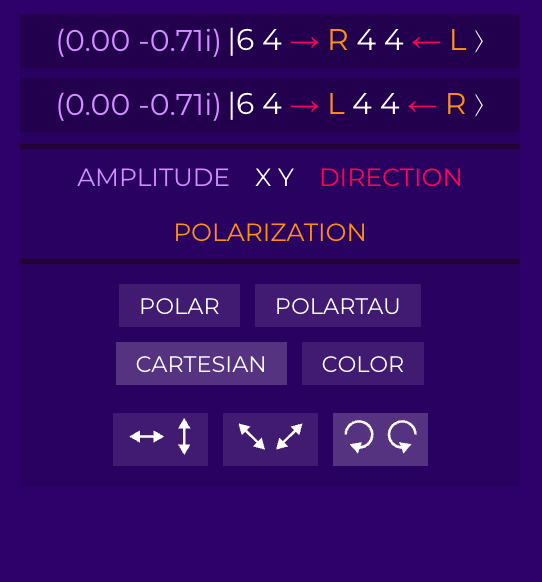}
\end{tabular}
\end{center}
\caption 
{ \label{fig:braketvue_state_vl}
A Bell state $\ket{\Phi^+}$ represented as a ket vector, with various choices of bases and complex number formatting. In this notation, $\ket{6\ 4\rightarrow H}$ is a single-photon superposition component located at position $(6, 4)$, travelling right, with horizontal polarization; $\ket{6\ 4\rightarrow H\ 4\ 4\leftarrow H}$ is a two-photon state.} 
\end{figure} 

Users can check the exact quantum state at a given simulation step in the typical form of kets\cite{dirac_new_1939}. We provide a few improvements for readability and interactivity, so to create a clear and minimalistic visual data representation\cite{brinton_graphic_1939,tufte_visual_2001}, see Fig.~\ref{fig:braketvue_state_vl}. We employ color-coding so to match the formula with its description\cite{riffle_understanding_2011,euclid_first_1847,rougeux_byrnes_2018}. Users can choose between various representations of complex numbers -- polar, polar using full rotations ($\tau=2\pi$), Cartesian, and a color circle (amplitude as radius, phase as hue). Users can change between these representations according to their preferences or what is most suitable for a given phenomenon.

Likewise, we provide a way to dynamically change between bases -- horizontal ($\ket{H}$, $\ket{V}$), diagonal ($\ket{D}$, $\ket{A}$), and circular ($\ket{L}$, $\ket{R}$). This feature can be used to set a basis that works best for a given phenomenon or to toggle between bases to explore the same phenomena in a few different ways. For example, the Faraday rotator, when expressed in $\ket{H}$ and $\ket{V}$, acts as a rotation. But for the circular basis of $\ket{L}$ and $\ket{R}$, it is just a phase change between polarizations. Likewise, for a mirror in the horizontal basis, due to changing the coordinate system, horizontal polarization gets a sign flip, which might look unremarkable. Yet, in the circular basis, we see that this sign flip swaps left with right circular polarization.

\begin{figure}
\begin{center}
\includegraphics[width=0.85\textwidth]{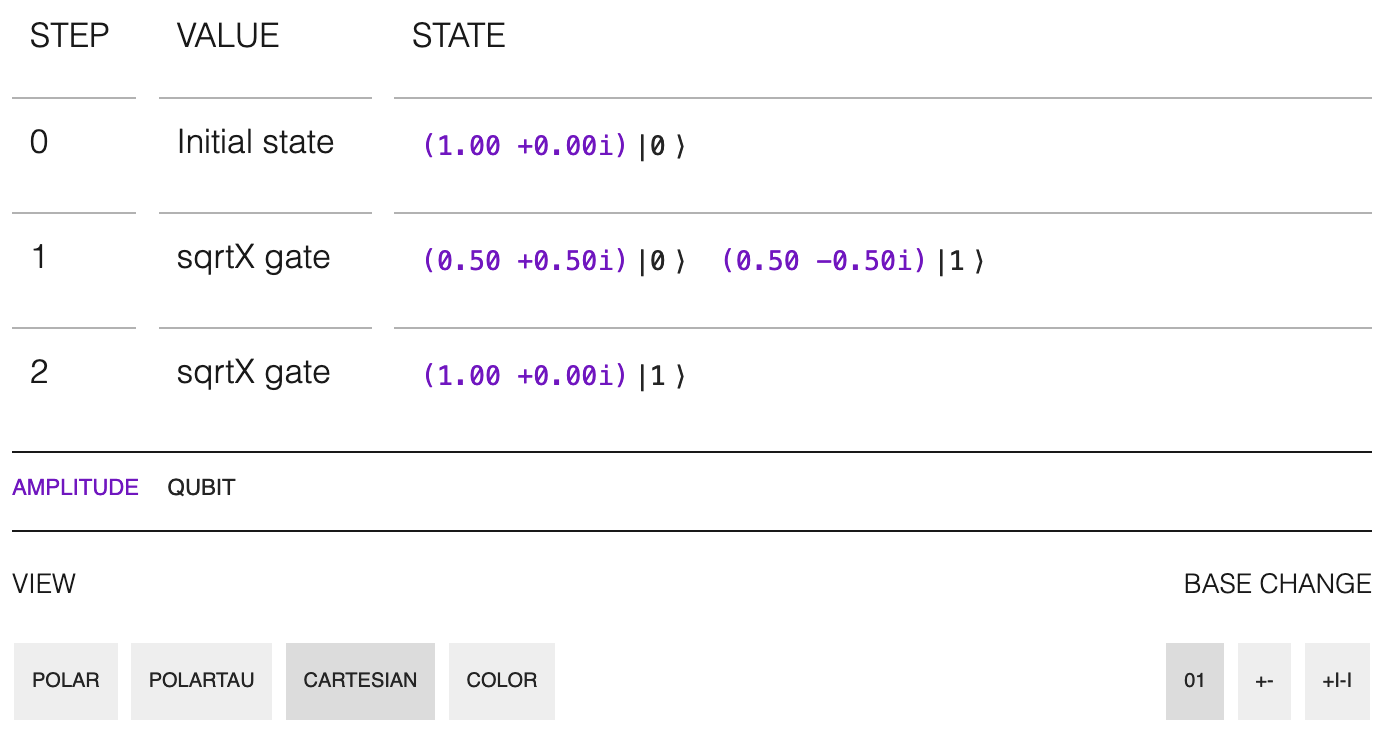}
\end{center}
\caption 
{ \label{fig:braketvue_state_qubit}
A different example of the \emph{BraKetVue} state visualizer, a single-qubit state propagating through a quantum circuit. } 
\end{figure}

We distribute an extended version of this feature as an open-source library \emph{BraKetVue}\cite{jankiewicz_braketvue_2022}, also available at \url{https://github.com/Quantum-Flytrap/bra-ket-vue}.
This public version supports arbitrary discrete states (spin, qubits, n-energy states, and user-defined), sequences of states, and light \& dark mode, see Fig.~\ref{fig:braketvue_state_qubit}. Since \emph{BraKetVue} is a \emph{JavaScript} library, it can be used on websites, in interactive materials, on web-based slides, or embedded in \emph{Jupyter Notebook}. One of the examples is an explorable explanation of quantum logic gates for a single qubit\cite{zendejas-morales_quantum_2021}, \url{https://quantumflytrap.com/blog/2021/qubit-interactively/}.

\subsection{Operator visualization}
\label{ss:operator}

One typical way to visualize quantum operators involves showing them in a braket notation, or as any other matrix -- displaying them as a table with numerical values. This approach is insufficient for clear presentation.

In data visualization, arrays can be visualized with heatmaps -- two-dimensional tile plots, with tile color representing the numerical value. For an interactive example, see \emph{Clustergrammer}\cite{fernandez_clustergrammer_2017}. However, typical plotting packages (such as \emph{Python Matplotlib} or \emph{R ggplot2}) do not support complex numbers.

\begin{figure}
\begin{center}
\includegraphics[width=0.6\textwidth]{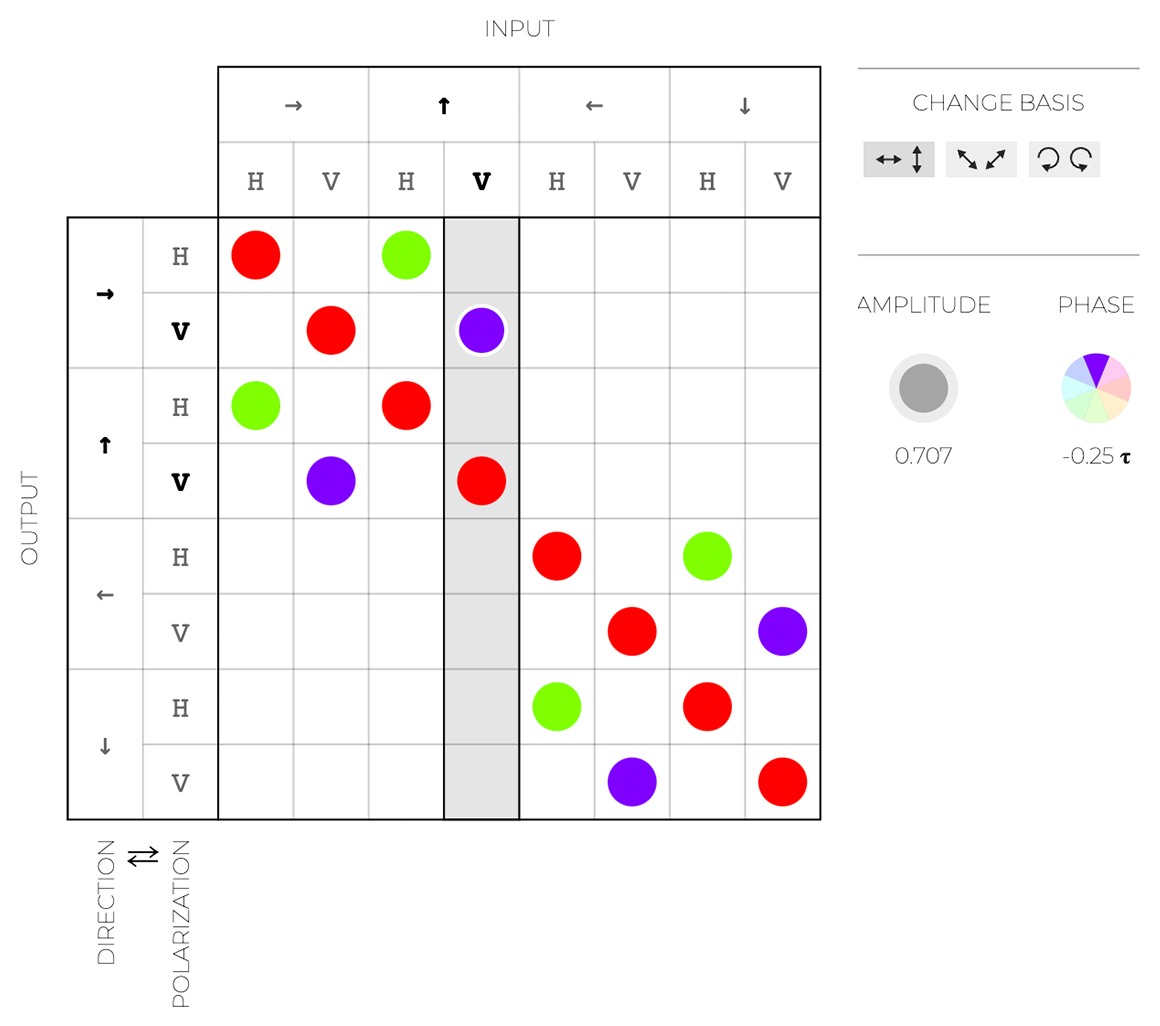}\\
\includegraphics[width=0.6\textwidth]{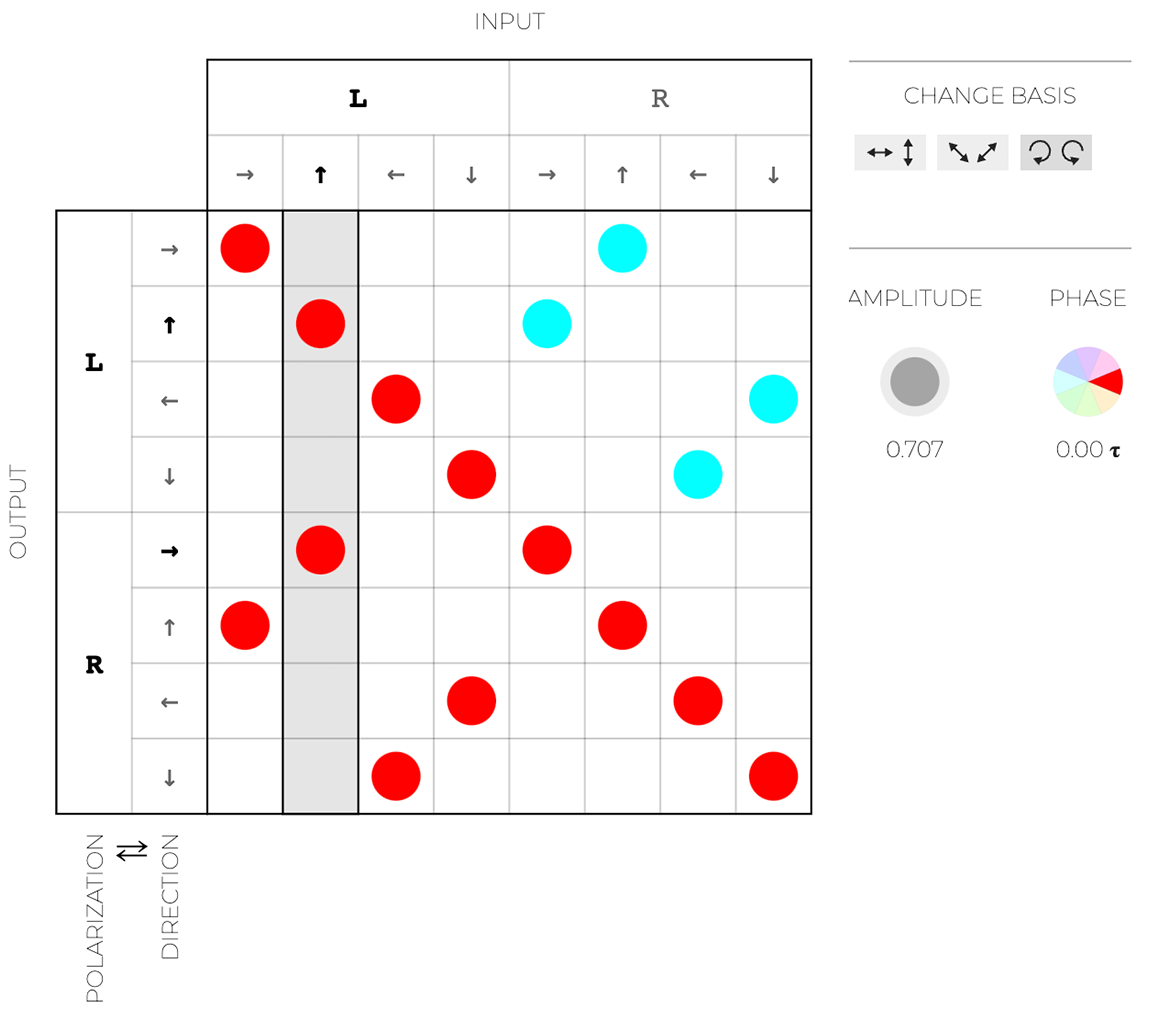}
\end{center}
\caption 
{ \label{fig:braketvue_operator_bs}
\emph{BraKetVue} visualizations of matrices representing quantum operators for a non-polarizing beamsplitter. The interactive legend show $r$ and $\phi$ of highlighted matrix entry. Top: the horizontal/vertical basis, direction-polarization tensor ordering. Bottom: the circular basis, polarization-direction tensor ordering.}
\end{figure}

\begin{figure}
\begin{center}

\begin{tabular}{cc}
    \includegraphics[width=0.4\textwidth]{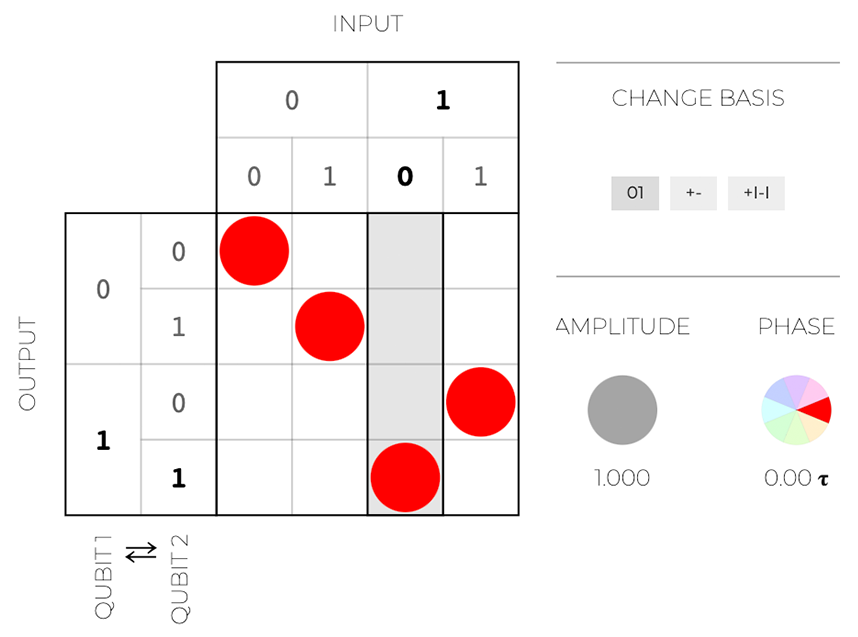} & \includegraphics[width=0.4\textwidth]{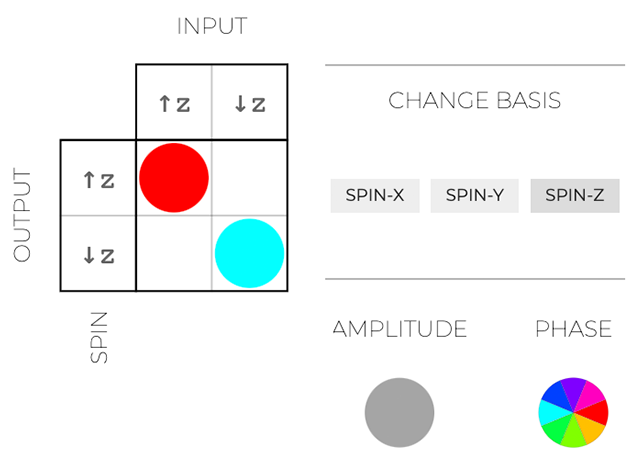}
\end{tabular}
\end{center}
\caption 
{ \label{fig:braketvue_operator_other}
\emph{BraKetVue} visualizations of matrices beyond \emph{Virtual Lab}. Left: the Controlled-NOT (CNOT) quantum gate. Note that with the input column highlighted, we get a truth table; in this case, we see that: $\ket{10} \mapsto \ket{11}$.
Right: the Pauli-Z operator for a spin-1/2 system.}
\end{figure}

We present a quantum operator matrix as an array of colored disks, with basis state labels, and with further interaction options for label rearrangement and selection of basis,  Fig.~\ref{fig:braketvue_operator_bs} and Fig.~\ref{fig:braketvue_operator_other}.
We followed a similar visualization scheme to a recursive pattern of \emph{Qubism}\cite{rodriguez-laguna_qubism_2012,migdal_symmetries_2014}, applied to operators (matrices) instead of wavefunctions (complex vectors). For a complex amplitude $z=r \exp(i \phi)$, we represent its length $r$ as a disk radius and its phase  $\phi$ as hue. The color scheme follows a domain coloring technique\cite{farris_domain_2017}.  Instead of the common choice to map $r$ to brightness or opacity, we use disc radius for a few reasons. First, it is easier to read regardless of screen settings, even for low amplitudes. Second, the disk area $\pi r^2$ is proportional to another vital property -- probability. 
We show the amplitude and phase values on mouseover, with three applications in mind: as a chart legend, as access to the exact values, and provide accessibility for people with color blindness (regardless of its type).

\subsection{Visualizing entangled amplitudes}
\label{ss:amp_entanglement}

\begin{figure}
\begin{center}
\begin{tabular}{ccc}
\includegraphics[height=0.1\textwidth]{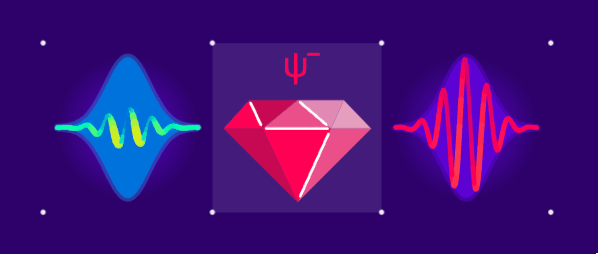} &
\includegraphics[height=0.1\textwidth]{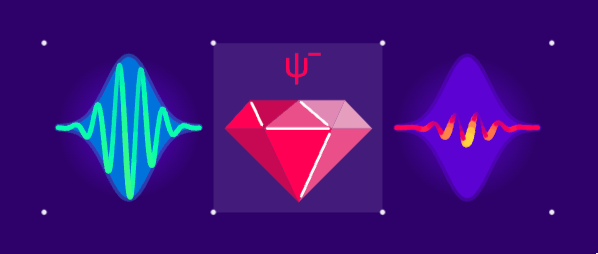} &
\includegraphics[height=0.1\textwidth]{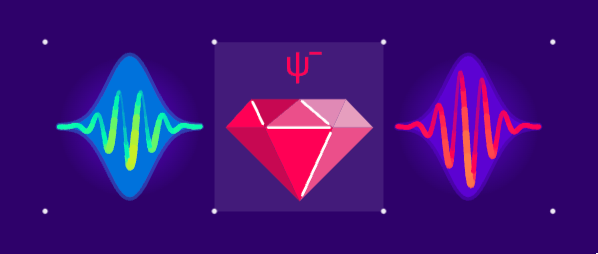}
\end{tabular}
\end{center}
\caption 
{ \label{fig:entangled_amplitudes}
Three snapshots of the amplitude blinking visualizing entanglement in the singlet state $\ket{\Psi^-}$. Note that within an image, the states are orthogonal to each other.} 
\end{figure} 

To visualize entanglement, we invented a new method since we didn’t find any procedural way of visualizing arbitrary entangled states. Visually, we represent entanglement by coordinated blinking of the entangled particles, see Fig.~\ref{fig:entangled_amplitudes}. For two particles in state $\ket{\Psi}_{12}$, we use the following procedure:
\begin{enumerate}
    \item We take a random one-particle state $\ket{\chi_2}$ from a uniform distribution.
    \item We get $\ket{\psi_1} = \bra{\chi_2}_2 \ket{\Psi}_{12}$ for the first particle.
    \item We get $\ket{\psi_2} = \bra{\psi_1}_1 \ket{\Psi}_{12}$ for the second particle.
\end{enumerate}

That is, we sample the state of of the first particle $\ket{\psi_1}$ if the other particle is randomly measured. Then we get the corresponding state of the second particle $\ket{\psi_2}$. We get two states, $\ket{\psi_1}$ and $\ket{\psi_2}$, which display separately the same way. For an easy example, let’s take Bell state $\ket{\Psi^-} = (\ket{01} - \ket{10}) / \sqrt{2}$, also known as the singlet state. In this state, two particles are always orthogonal. Let's check that:
\begin{align}
    \ket{\chi_2} &= \cos(\alpha) \ket{0} + \sin(\alpha) \exp(i \phi) \ket{1}\\
    \ket{\psi_1} &= \frac{- \sin(\alpha) \exp(-i \phi) \ket{0} + \cos(\alpha) \ket{1}}{\sqrt{2}}\\
    \ket{\psi_2} &= -\frac{\cos(\alpha) \ket{0} + \sin(\alpha) \exp(i \phi) \ket{1}}{2}
\end{align}
So indeed, up to a normalization for $\ket{\Psi^-}$, we get orthogonal states $\ket{\psi_1}$ and $\ket{\psi_2}$.

We can easily generalize this operation for $n$ particles. First, we generate $n-1$ random vectors $\ket{\chi_2}, \ldots \ket{\chi_n}$. Taking a random unit vector for $d$-dimensional space is simple and does not require parameterization. We generate $2d$ random numbers with Gaussian distribution\cite{box_note_1958} (for the real and imaginary parts of each coordinate) and then normalize the vector. We do not need to generate all random vectors (i.e. around 1000 per photon) -- it suffices to generate a subset for which photon probabilities are non-zero. Then, we follow an iterative procedure:
\begin{align}
    \ket{\psi_1} &= \bra{\chi_2}_2 \ldots \bra{\chi_n}_n \ket{\Psi}\\
    \ket{\psi_2} &= \bra{\psi_1}_1 \bra{\chi_2}_3 \ldots \bra{\chi_n}_n \ket{\Psi}\\
    &\cdots\\
    \ket{\psi_n} &= \bra{\psi_1}_1 \ldots \bra{\psi}_{n-1} \ket{\Psi}
\end{align}
In particular, for $n=1$ we recover exactly the initial state. The ordering of particles does influence the result. However, for an interactive, blinking visualization we have found it is a non-issue.

An alternative approach we considered is the Schmidt decomposition of an entangled state:
\begin{equation}
    \ket{\Psi} = \sum_i \lambda_i \ket{\psi_1^i} \ket{\psi_2^i}
\end{equation}
However, this method would cause some issues. First, we would need to perform a costly operation Singular Value Decomposition (SVD) at each state. Second, if some $\lambda_i$ are the same, this decomposition is not unique (e.g. for Bell pairs). Hence, it is not stable -- extremely small changes in the state (including due to using floating-point numbers) will result in a non-continuous change in the visualization. Third, the Schmidt decomposition is not easily generalizable to states of three particles. While it is possible to hierarchically decompose it into two steps: 1-(23) and then 2-3, this depends on the order of the photons.

Both approaches (ours and the Schmidt decomposition) can be simply applied to the visualization of any quantum state, including qubits, electrons, and other systems. In general, it is impossible to present entangled quantum states as probabilistic representations.  Consequently, this coordinated blinking should be understood as a quantum correlation rather than a statistical mixture.

\subsection{Entanglement between particles}
\label{ss:entanglement_graph}

\begin{figure}
\begin{center}
\begin{tabular}{cc}
\includegraphics[height=0.2\textwidth]{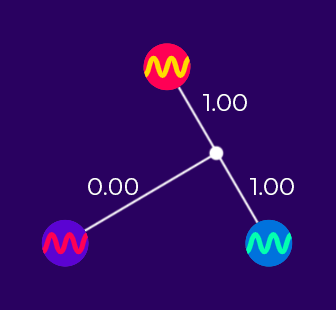} &
\includegraphics[height=0.2\textwidth]{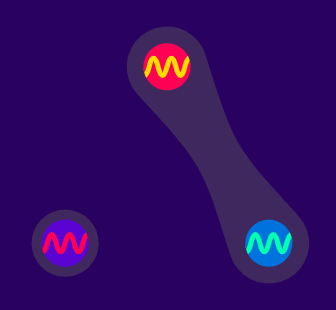}
\end{tabular}
\end{center}
\caption 
{ \label{fig:entangled_measures}
An entanglement graph for three particles, for a bipartite system of a photon and an entangled Bell pair. Left: the spring model, with an entanglement measure serving as the spring coefficients. The connecting point is the equilibrium. Right: the same graph visualized as a blob connecting entangled particles --- each connection width is a function of the entanglement measure.} 
\end{figure} 

While the blinking of the waves is an indicator of entanglement, it might be hard to see the degree of entanglement between parts of the system. Since we didn’t find an existing way to represent the degree of entanglement between particles, we created our own. Working with pure states simplifies the task, as all measures of correlations are related to entanglement.

A quantum physics novice might expect to see values of entanglement between each particle pair. That is, for three particles A, B, and C one would expect to have some measures of entanglement for A-B, B-C, and C-A. However, it is not possible. The only meaningful measure is the entanglement of a particle with the rest of the system: A-BC, B-AC, C-AB.
We use Rényi-2 entanglement entropy, due to its computational simplicity.
\begin{equation}
    H_{A,BC} = - \log_2 \text{Tr} \left[  \text{Tr}_{AB} [ \projection{\Psi} ]^2  \right]
\end{equation}

Next, we turn this measure into a digestible visualization, see Fig.~\ref{fig:entangled_measures}. 
We place particles in a circle and represent them as dots. We draw a string connecting them and set the respective entanglement entropy (of a particle with the rest of the system) as the force constant. If all strings are equally stiff, their connecting points land in the middle. If only two particles are entangled with each other, then the connecting point lands between them. Next, we turn the strings into amoeba-blobs (force constants become connection widths) so that we can clearly see which particles are entangled with each other. The same approach can be employed to visualize entanglement in any other system of quantum particles.

\subsection{Multiverse tree}

\begin{figure}
\begin{center}
\includegraphics[height=0.4\textwidth]{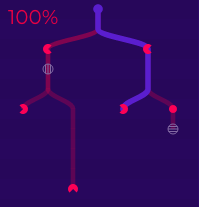}
\end{center}
\caption 
{ \label{fig:multiverse_tree}
 A multiverse tree with a highlighted path to the state we currently explore. Each absorption event is represented by an icon of the absorbing element, including polarizers and detectors. As the tree can be infinite for setup with loops, we show the percentage of all simulated paths, weighted by their probabilities.} 
\end{figure} 

We represent the evolution of quantum states as a tree rather than a sequence of steps, see Fig.~\ref{fig:multiverse_tree}. Each measurement event splits evolution into as many branches as there are outcomes with non-zero probability. The multiverse tree serves as a summary of all possible outcomes and as a way to explore the experiment. The user can click to select each timestep of each branch.

There are several interpretations of quantum mechanics, including the indeterministic Copenhagen interpretation and the deterministic Everettian many-worlds interpretation\cite{zurek_decoherence_2003}. Even among physicists, the choice of interpretation is a matter of philosophical belief rather than a scientific hypothesis\cite{schlosshauer_snapshot_2013,sivasundaram_surveying_2016}. The multiverse tree works the same regardless of the interpretation, just with a difference in what the nodes mean -- separate probabilistic branching events or components of the global wavefunction.

\subsection{Classical randomness, table of measurements, and correlators}

\begin{figure}
\begin{center}
\includegraphics[height=0.4\textwidth]{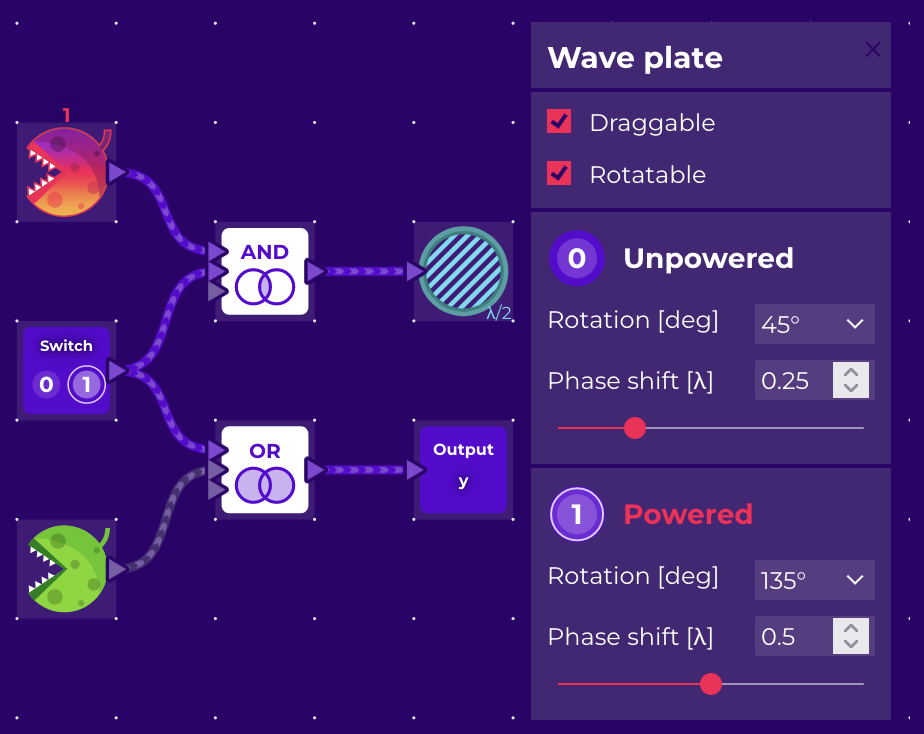}
\end{center}
\caption 
{ \label{fig:wires}
Inputs connected through logic gates (AND and OR) to an element and an output variable. A gray wire carries $0$ (false) and a highlighted $1$ (true). An element connected to a value has two sets of adjustable parameters.} 
\end{figure}

In addition to quantum states, \emph{Virtual Lab} supports classical states based on the initial setup and detection events. They are transferred through wires, with logic operations, and used for modifying elements on the board, measuring output correlations, and logging the results for arbitrary analysis, see Fig.~\ref{fig:wires}. The data are downloadable as a CSV file. For a full example, see the Bell inequality violation in Fig.~\ref{fig:experiment_bell_inequality}.

Input variables can be deterministic (so work as switches) or randomized (so on each run are randomly set to on/off). Both have their application in setups -- especially in quantum key distribution protocols and other correlation-based experiments. Since elements are affected by detection events, it is possible to create conditional setups, e.g. quantum teleportation.

All wires carry their signals instantly, without the speed-of-light delay, so to simplify the interface.

Correlators measure correlation between $\pm 1$-valued variables $s_1, s_2, \ldots s_n$, conditional on the control variables $c_1, c_2, \ldots c_m$. For a practival example, see the Bell inequality in Fig.~\ref{fig:experiment_bell_inequality}.

\subsection{Quantum Game}

The sandbox mode offers a way for the user to create arbitrary setups. However, sandbox mode by design lack any objectives and can be intimidating as it provides all functionalities at once. It is entirely up to the user to decide how to interact with \emph{Virtual Lab}.

\begin{figure}
\begin{center}
\includegraphics[width=0.85\textwidth]{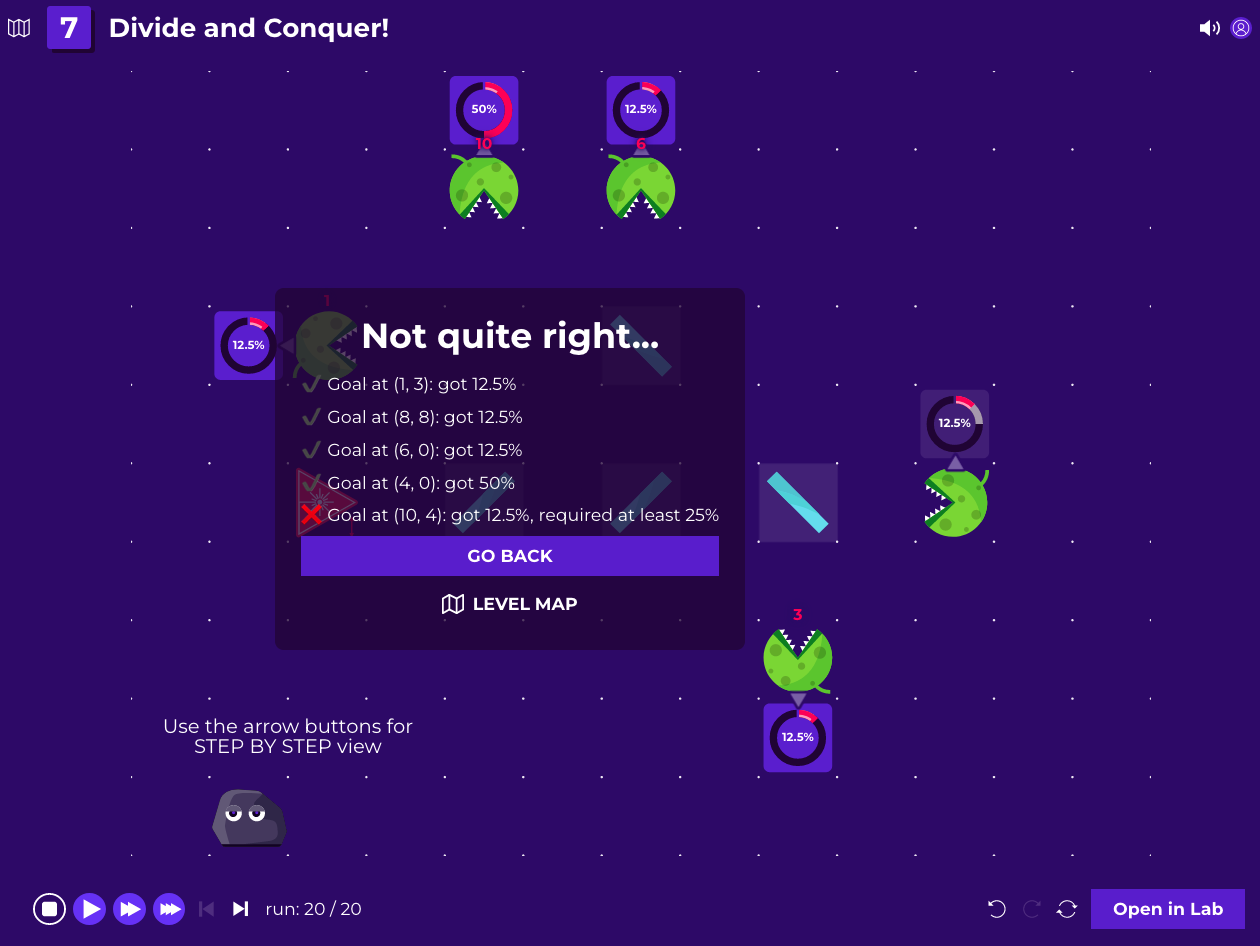}
\end{center}
\caption 
{ \label{fig:game}
A screen from \emph{Quantum Game} after running a simulation shows level goals and their status.} 
\end{figure}

Based on the earlier success of \emph{Quantum Game with Photons}, we created a similar experience, see Fig.~\ref{fig:game}. Rather than releasing a separate product, we made \emph{Quantum Game} a part of \emph{Virtual Lab}. While our focus was to provide an enjoyable experience for beginners and advanced users alike, we wanted it to serve as a step-by-step tutorial to both quantum physics and the \emph{Virtual Lab} interface. We aimed to make it clear that \emph{Quantum Game} is an advanced simulation of quantum mechanics, rather than a visual gimmick loosely based on actual physics.

For each level, the gameplay consists of creating a setup, running it many times, and collecting the results -- a typical pattern for programming games\cite{blanco_patterns_2020}, e.g. \emph{SpaceChem} and \emph{Shenzhen I/O} by \emph{Zachtronics}. The probabilistic nature of quantum mechanics makes it challenging to design clear goals. We opted for setting detection percentage thresholds (e.g. a given detector needs to absorb a photon with the probability greater or equal to $25\%$). While we show a finite number of experimental runs (e.g. 20), the win condition is based on probabilities in the infinite limit.

\section{Simulation}
\label{sec:simulation}

\emph{Virtual Lab} simulates quantum systems by using the pure state of a fixed number of particles, a discrete basis, grid-based positions, discrete-time evolution, and exploration of all measurement outcomes.
We selected these constraints in order to create a fast, robust simulation of typical quantum experiments, without adding extra complexity (both numerical and related to the user experience). We have found a sweet spot -- suitable for a \emph{LEGO bricks}-like interface, general enough to simulate major quantum information experiments, viable numerically, and directly applicable to simulating quantum devices.

The whole simulation is stored as a tree of classical and quantum states, explorable with the multiverse tree interface. Each node contains its probability, the quantum state, and the classical state.
We open-sourced our first numeric engine written in \emph{TypeScript}, \emph{Quantum Tensors}\cite{migdal_quantum_2022-2}, available at \url{https://github.com/Quantum-Flytrap/quantum-tensors}.
It serves as a reference for the simulation steps and is ready for use both in \emph{Virtual Lab} and for arbitrary quantum systems with discrete states -- e.g. quantum circuits, spin, and qudits. We put care into creating a semantic tensor structure\cite{chiang_named_2021}, for the clarity and safety of advanced operations on complex vectors and matrices.

\subsection{State representation}

The state of a single photon is characterized by four components: position (along the x and y axes), direction, and polarization. A photon has two polarization states. Since we use grid mechanics, light travels along four cardinal directions, rather than at an arbitrary angle. Grid size is a parameter that can be adjusted according to needs, and we use the default of $13\times10$, which we find suitable for most applications. It gives rise to around 1000 dimensions per single particle.
\begin{equation}
    \ket{\psi} = \sum_{x,y,dir,pol} \alpha_{x,y,dir,pol} \ket{x}\ket{y}\ket{dir}\ket{pol}
\end{equation}

\emph{Virtual Lab} simulation supports primarily distinguishable particles, while photons are indistinguishable bosons. However, they can be made distinguishable by assigning them a wavelength. A non-entangled product state is represented as a tensor product of single-particle states:
\begin{equation}
    \ket{\Psi} = \ket{\psi_1}\ket{\psi_2}\ket{\psi_3}
\end{equation}
We simulate indistinguishable photons (e.g. for the Hong-Ou-Mandel effect\cite{hong_measurement_1987}) via symmetrization:
\begin{equation}
    \ket{\Psi} = \frac{\ket{\psi_1}\ket{\psi_2} + \ket{\psi_2}\ket{\psi_1}}{2 + 2 \left| \braket{\psi_1}{\psi_2} \right|^2}
\end{equation}

This approach is equivalent to the typical description of bosons with creation and annihilation operators for states with a fixed number of photons\cite{migdal_which_2014,migdal_symmetries_2014}. However, from a numerical point of view, it is easier to work with vectors than polynomials. One significant drawback of our choice is the restriction to Fock states -- the current framework does not support coherent or squeezed states.

\subsection{Unitary operations}

In each step, we perform three operations: free propagation, measurement, and the action of unitary operation. The propagation step $T$ involves shifting the photon’s position according to its direction:
\begin{equation}
    T = \sum_{x, y, pol, dir} \ket{x + dir_x}\bra{x} \otimes \ket{y + dir_y}\bra{y} \otimes \projection{dir} \otimes \projection{pol}
\end{equation}
Unitary single-photon operations (i.e. all passive optics) are described by:
\begin{equation}
    \ket{\Psi(t+1)} = U^{\otimes n} \ket{\Psi(t)}
\end{equation}
For an $n$-photon state, $U^{\otimes n}$ is a tensor power, e.g. $U^{\otimes 3} = U \otimes U \otimes U$. The non-signaling principle from special relativity forbids instant interactions between spatially-separated particles and thus the operator can be decomposed as
\begin{equation}
    U = \sum_{x,y} \projection{x, y} \otimes U_{x,y}
\end{equation}
where operator $U_{x,y}$ corresponds to the element on the board. The operator $U$ may change in time if it is connected to a classical wire.

We use a limited number of two-photon interactions. This choice is motivated by optical setups, where photon-photon interactions are challenging to implement. We apply this operation:
\begin{equation}
    \ket{\Psi(t+1)} = V_{12} V_{21} \ket{\Psi(t)}
\end{equation}
where $V_{21}$ is the tensor transpose of $V_{12}$, i.e. the same operator but with swapped photon order. We ensure that $V_{12}$ and $V_{21}$ commute. In particular, for a CNOT operator, we encode qubits in polarization states, use top-down direction ($\ket{\uparrow}$ and $\ket{\downarrow}$) for control, and left-right direction ($\ket{\leftarrow}$ and $\ket{\rightarrow}$ for the target. As all interactions are local, both photons need to arrive at the CNOT element at the same time. That is, the operator is:

\begin{align}
    P_{\updownarrow \leftrightarrow} &= \left( \projection{\uparrow} + \projection{\downarrow} \right) \otimes \left( \projection{\leftarrow} + \projection{\rightarrow} \right)\\
    CNOT &= P_{\updownarrow \leftrightarrow}  \otimes \left( \projection{HH} + \projection{HV} + \ket{VV}\bra{VH} + \ket{VV}\bra{VH} \right)\\
    &+ \left( I_{dir} \otimes I_{dir} - P_{\updownarrow \leftrightarrow}  \right)  \otimes I_{pol} \otimes I_{pol}
\end{align}
Since $P_{\updownarrow \leftrightarrow}$ and $P_{ \leftrightarrow \updownarrow}$ are projections on orthogonal spaces, $CNOT_{12}$ commutes with $CNOT_{21}$.

\subsection{Measurement}

We allow measurements at any step, including destructive measurements (absorbing a photon) and nondemolition measurements\cite{braginsky_quantum_1980} (not absorbing a photon). In each measurement phase and for each photon there are three exclusive outcomes:
\begin{itemize}
    \item The photon got measured and destroyed.
    \item The photon got measured but not destroyed.
    \item The photon didn’t get detected.
\end{itemize}
Note that after a measurement event, the state remains a state with a fixed number of photons.

All measurements happen within the elements on the board. Destructive measurements happen when a photon hits a fully absorptive element (e.g. a rock, a detector, the side of certain elements), a partially absorptive element (e.g. a neutral density filter), or a polarization-dependent element (e.g. linear polarizer), or falls out of the board. Nondemolition measurements happen when a photon passes through a nondemolition detector or when none of the other detection events occur.

Destructive measurements are represented with a non-normalized (bra) vector, $\sqrt{w_i} \bra{\phi_i}$. For example, for a single direction $\ket{\rightarrow}$, these vectors for polarization are:
\begin{itemize}
    \item Detector and rock: $\bra{H}$ and $\bra{V}$.
    \item Neutral-density filter: $\sqrt{a} \bra{H}$ and $\sqrt{a} \bra{V}$, where $a$ is the absorption rate.
    \item Linear polarizer: $\cos(\alpha) \bra{H} + \sin(\alpha) \bra{V}$, where $\alpha$ is the polarizer rotation.
\end{itemize}
After a measurement of the $m$-th photon, the remaining state is
\begin{equation}
    \sqrt{w_i} \bra{\phi_i}_m \ket{\Psi}
\end{equation}
and has one less photon. The probability of the outcome is this state's norm squared.

To deal with the general case of destructive and nondemolition measurements, we need to use a general scheme of Positive Operator-Valued Measures (POVMs). A general POVM requires a set of positive semi-definite matrices $M_i$ that sum up to identity\cite{nielsen_quantum_2010}. The resulting non-normalized state for three photons is:
\begin{equation}
    \ket{\Phi(i,j,k)} = \left( \sqrt{M_i} \otimes \sqrt{M_j} \otimes \sqrt{M_k} \right) \ket{\Phi}
\end{equation}
with its norm squared being the probability of the given outcome. Here we understand $\sqrt{M}$ as an operator such that $\sqrt{M}^\dagger \sqrt{M} = M$.
In our simulation, we restrict ourselves to $M_i$s that form a set of orthogonal, weighted projections:
\begin{align}
    M_i &= w_i P_i\label{eq:scaled_proj}\\
    P_i P_j &= 0 \text{ for all } i \neq j
\end{align}
This simplifies calculating the new state as we do not need to take the square root of an operator, which is numerically costly and not uniquely defined. For scaled projective measurements the root is $\sqrt{w_i} P_i$, up to a unitary operator acting on its subspace, which we arbitrarily set to identity. While this is a restriction, in the case of \emph{Virtual Lab}, we didn’t find a practical case in which the assumption \eqref{eq:scaled_proj} is an issue. On our board, some elements are absorptive, some perform the nondemolition measurement, and there is the null result of non-detection. The latter is represented as:
\begin{equation}
    M_{\star} = I - \sum_i M_i
\end{equation}
All $M_i$ are orthogonal (as they are related to spatially different tiles in the simulation) and weights $w_i \leq 1$. Consequently, $M_{\star}$ is a positive semidefinite operator. With these assumptions, we can take its root:
\begin{equation}
    \sqrt{M_{\star}} = I + \sum_i \left( \sqrt{1 - w_i} - 1 \right) P_i
\end{equation}

\subsection{Performance}

To create a smooth experience, we needed to have fast response times. As a rule of thumb, up to a $100 ms$ response time feels like a system reacts instantaneously, while up to a $1s$ response time still keeps the user’s flow of thoughts uninterrupted\cite{nielsen_usability_1993, card_psychology_2008,nielsen_response_2014}. This sets the upper bound for the simulation time of up to three photons. For comparison, 60 frames per second required for a smooth video game experience\cite{claypool_frame_2007} corresponds to $17 ms$. None of the off-the-shelf libraries (\emph{NumPy}, \emph{SciPy}, \emph{TensorFlow}, \emph{PyTorch}) were able to support this kind of operation with these constraints, nevermind working in a browser.

Note that even for a three-photons state with a typical (dense) representation, $10^9$ complex numbers are required. A single unitary operator would require $10^{18}$. Yet, in typical experimental cases, we use only a few non-zero amplitudes -- prompting us to use sparse vectors.  This task goes far beyond typical interactive simulators of quantum circuits. For 8 qubits, it is $2^8 \approx  10^3$ dimensions, which can be easily simulated with dense vector and matrix operations.

However, even with sparse vectors for 3 particles, an identity matrix would still have $10^9$ non-zero elements (i.e. the ones on the diagonal). Hence, we need to apply operations to take advantage of the structure of the tensor, for speedup of operators acting only on a subset of dimensions. 

The initial approaches are in an open-source \emph{TypeScript} library \emph{Quantum Tensors}. To improve performance even further, we developed a high-performance sparse array simulation in \emph{Rust}, a low-level programming language. The typical timescale of simulating experiments is $30-150 \mu s$. To use the engine in the browser, we compiled \emph{Rust} code to \emph{WebAssembly} (\emph{WASM})\cite{haas_bringing_2017}. We measured that the execution time in the browser is only $1.5\times$ slower than native code.

\section{Example experiments}
\label{sec:experiments}

\emph{Virtual Lab} lets users simulate the physics of quantum information, quantum computing, and quantum cryptography\cite{kok_review_2007}, as well as classical fields of wave optics (including polarization and interference) and electrodynamics.

We group the experiments by the number of photons involved. A fair share of the single-photon experiments are directly related to the behavior of classical light\cite{waseem_quantum_2020} -- even though a laser emits coherent states, rather than a series of single photons. Hence, we make a distinction between classical and quantum one-photon experiments. The list we provide is not exhaustive due to the open-ended nature of the experiments. All of the following experiments can be simulated in \emph{Virtual Lab}, see
\url{https://lab.quantumflytrap.com/experiments}.

We suggest this list as a starting point of didactic material for courses, as an interactive lab used together or in place of physics experiments, or for a more scalable experience available to a larger number of participants. 

\subsection{One photon, classical}

Interference is a wave phenomenon in which waves add by their amplitudes and amplify or cancel each other out depending on their relative shift. This behavior might be counterintuitive to people accustomed to geometric optics, in which light adds by its intensity.

In optics, interferometers are applied for sensitive measurements of distance. The famous experiment of \textbf{Michelson and Morley}\cite{michelson_relative_1887} in 1887 showed that light velocity is independent of the frame of reference, thus setting footing for the special relativity theory. For an accessible introduction to special relativity, we suggest \textit{"Unusually Special Relativity"} by Andrzej Dragan\cite{dragan_unusually_2021}. For an interactive version of the experiment, see
\url{https://lab.quantumflytrap.com/lab/michelson-morley}.

\begin{figure}
\begin{center}
\includegraphics[width=0.6\textwidth]{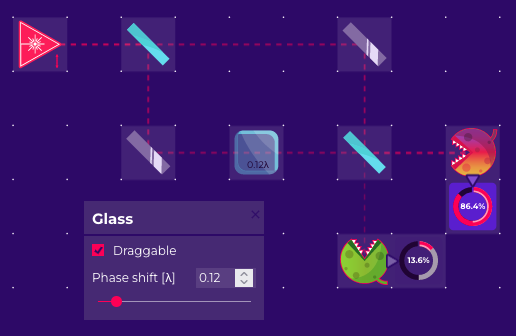}
\end{center}
\caption 
{ \label{fig:experiment_mach_zehnder}
Changing relative optical paths in the Mach-Zehnder interferometer. We use the high-intensity visualization mode.} 
\end{figure} 

The \textbf{Mach-Zehnder interferometer} (\url{https://lab.quantumflytrap.com/lab/mach-zehnder}) is applied in precise measurements of tiny changes of refraction index and for a wide range of quantum experiments. In the interactive version, the user can modify the relative delay between paths and see the result in real-time, see Fig.~\ref{fig:experiment_mach_zehnder}.

The \textbf{Sagnac interferometer}\cite{sagnac_lether_1913} (\url{https://lab.quantumflytrap.com/lab/sagnac-interferometer}), originally created to detect the luminiferous aether (as Michelson-Morley), is now used in ring laser gyroscopes and fiber optics gyroscopes.

Another set of phenomena involves the polarization of light. Understanding polarization is important for learning the nature of light and its application in numerous fields including photography, LCD displays, and telecommunications.
\textbf{The three polarizer paradox} (\url{https://lab.quantumflytrap.com/lab/three-polarizer-paradox}) is an experiment in which two perpendicular polarizers block light, but inserting a third allows some of the light to pass. This represents a classical version of the quantum Zeno experiment.

\textbf{Optical activity} is a phenomenon in which chiral molecules (e.g. D-glucose, a naturally occurring enantiomer) rotate the polarization of light. While in quantum optical laboratories this effect is not being applied to rotate polarization, it is being utilized to measure the concentration of sugar.
The \textbf{Faraday effect} is a magneto-optic effect where the polarization of light rotates in the presence of a magnetic field in a transparent material.
While this rotation may seem the same as the one caused by optical activity, it has different symmetries. That is, the rotation is opposite depending on if the light propagates in the same or opposite direction of the magnetic field. The effect is used in  virtually all other optical elements that are not symmetric in time, such as optical circulators and optical diodes, see \url{https://lab.quantumflytrap.com/lab/optical-diode}. 

\subsection{One photon, quantum}

There is a common misunderstanding that at least two entangled particles are required to get distinctively quantum phenomena. Even some parts of quantum computing are possible without entanglement\cite{lloyd_quantum_1999,lanyon_experimental_2008}.
First and foremost, in quantum mechanics, light is absorbed in discrete portions (or quanta), with energy $\hbar \omega$ each. While at classical intensities of light we can measure the power of a light beam (or the energy of a pulse), for low intensities the detection phenomenon has a discrete nature.

Even the detection of a single particle serves as a way to explore the fundamentals of quantum mechanics, such as the shot noise (randomness of detection events), Born’s rule (the probability of measurement is $|\psi|^2$), and the nature of measurement. Shot noise can be seen on photographic film and digital sensors. Consumer smartphone cameras can be used to extract pure quantum randomness\cite{sanguinetti_quantum_2014}. We can simulate more advanced detection schemes, such as an optical implementation of optimal \textbf{non-orthogonal state discrimination}\cite{barnett_quantum_2008}, see \url{https://lab.quantumflytrap.com/lab/nonorthogonal-state-discrimination}.

\begin{figure}
\begin{center}
\begin{tabular}{cc}
\includegraphics[height=0.31\textwidth]{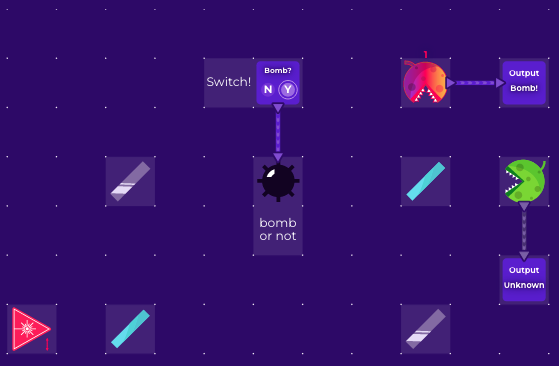} &
\includegraphics[height=0.31\textwidth]{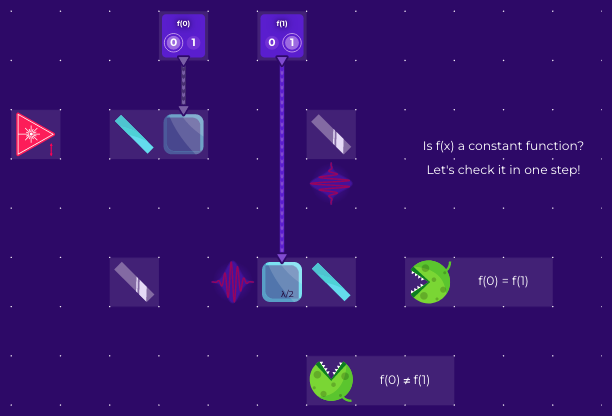}
\end{tabular}
\end{center}
\caption 
{ \label{fig:experiments_quantum_one_photon}
Left: The Elitzur–Vaidman bomb tester with a detection event that wouldn't be possible if the bomb were not present. Right: Deutsch-Jozsa algorithm for a function $\{0, 1 \} \rightarrow \{0, 1\}$ with its values $f(0)$ and $f(1)$ implemented as phase shifts. A single detection event tells us whether the functiom is constant.} 
\end{figure} 

\textbf{The Elitzur–Vaidman bomb tester}\cite{elitzur_quantum_1993,kwiat_experimental_1995} is a combination of interference and the nature of measurement. In the classical physics world, it is a tautology that a photo-sensitive bomb cannot be detected with light without setting it off. In the quantum world, there is room for “interaction-free measurement” by observing the changes in the interference pattern. A bomb can be placed in one arm of the Mach-Zehnder interferometer. Some detector clicks testify that there is something blocking one path and therefore breaking the interference, see Fig.~\ref{fig:experiments_quantum_one_photon} and
\url{https://lab.quantumflytrap.com/lab/elitzur-vaidman-bomb}.

Quantum measurement affects the measured quantum systems, even when no absorption is involved. It is impossible to measure two non-commuting observables without affecting them. Discovering where a particle is located destroys the interference pattern, which is called \textbf{quantum erasure}\cite{hillmer_-it-yourself_2007}, see \url{https://lab.quantumflytrap.com/lab/quantum-eraser}. Within \emph{Virtual Lab} is it not only possible to place a nondemolition measurement on one path, but also to define its efficiency, thus showing a continuous transition between fully breaking quantum interference and not touching the system at all, see \url{https://lab.quantumflytrap.com/lab/measurement-destroys-interference}.

The \textbf{Quantum Zeno effect} (\url{https://lab.quantumflytrap.com/lab/zeno-effect}) is another phenomenon in which a non-measurement affects the result. The state evolution is inhibited completely in the limit of infinitely frequent projective measurements.

Due to the fundamental rules of quantum mechanics, it is impossible to copy an arbitrary quantum state, also known as the \textit{“no-cloning theorem”}\cite{wootters_single_1982}. This property is a direct consequence of the linearity of quantum mechanics and is necessary for forbidding superluminal communication. This is a substantial challenge for quantum computing that provides an opportunity to use quantum states for cryptography. The \textbf{BB84 quantum key distribution protocol}\cite{bennett_quantum_2014} sends photons with random bits encoded quantum in different quantum bases -- any eavesdropping would be both imperfect (no-cloning) and discoverable (as measurement affects the system), see
\url{https://lab.quantumflytrap.com/lab/bb84}.

The \textbf{Deutsch-Jozsa algorithm}\cite{deutsch_rapid_1992, cleve_quantum_1998} (\url{https://lab.quantumflytrap.com/lab/deutsch-jozsa}) is an algorithm showing exponential speedup of a quantum algorithm over a classical one. It can be shown even within a single qubit that one needs a single measurement rather than two, see Fig.~\ref{fig:experiments_quantum_one_photon}.

Photon polarization serves as a carrier of quantum information. This dimension alone encodes a qubit. In \emph{Virtual Lab}, on top of optical operations, we provide abstract quantum gates. Last but not least, it is possible to encode \textbf{two classical qubits within a single photon}, using polarization for one dimension and direction for the other\cite{englert_universal_2001}. 

\subsection{Two photons}

It takes two particles to show one of the most striking quantum properties -- entanglement.

\textbf{Bell pairs} are a set of four quantum states forming a basis for two fully-entangled qubits. In \emph{Virtual Lab}, we provide an idealized non-linear crystal (based on spontaneous parametric down-conversion), which generates Bell pairs. Bell pair detection, necessary for some protocols (such as quantum teleportation), can be performed with a CNOT gate we provide. Experimental setups within \emph{Virtual Lab} can be applied to transform Bell states, explore correlation patterns, and create cryptography setups such as the \textbf{Ekert quantum key distribution protocol}\cite{ekert_quantum_1992}, see
\url{https://lab.quantumflytrap.com/lab/ekert-protocol}.

\begin{figure}
\begin{center}
\includegraphics[width=0.6\textwidth]{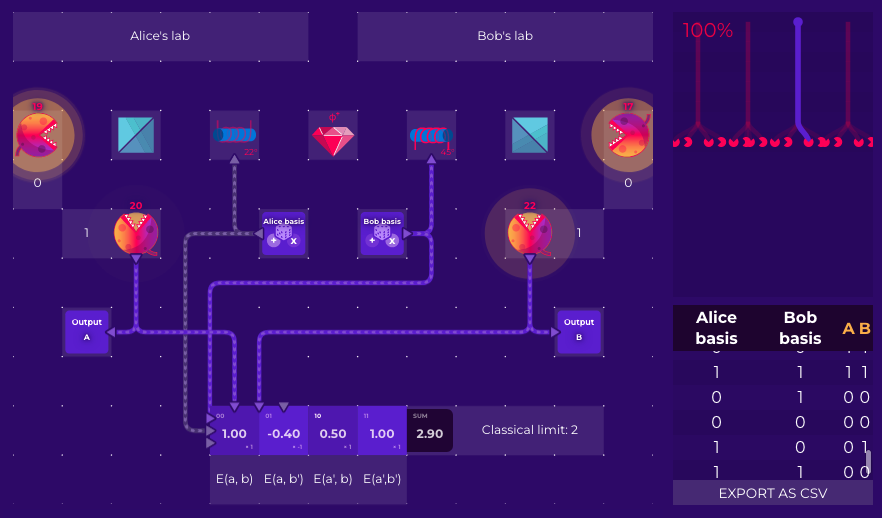}
\end{center}
\caption 
{ \label{fig:experiment_bell_inequality}
CHSH Bell inequality violation simulated with \emph{Virtual Lab}. We use the detection visualization mode.} 
\end{figure} 

The \textbf{Bell inequality}\cite{bell_einstein_1964} is a correlation that should hold for any probabilistic correlations between subsystems. It was created to turn a philosophical debate on hidden variables into an experimentally testable hypothesis. Quantum mechanics violates it with a Bell pair and a set of measurements. In \emph{Virtual Lab}, there is a correlator between results, set so that it can measure Bell inequality in the Clauser-Horne-Shimony-Holt (CHSH) form\cite{clauser_proposed_1969}. It can be used in two main ways -- as an exact result and as a quantity based on the number of detections (involving shot noise), see Fig.~\ref{fig:experiment_bell_inequality} and
\url{https://lab.quantumflytrap.com/lab/bell-inequality}.

\subsection{Three photons}

In \textbf{quantum teleportation}\cite{bennett_teleporting_1993}, a quantum state is transferred to another particle, no matter how far away. It requires three particles: one to be teleported and an entangled pair, with one particle being close to the source and the other in the target location. See Fig.~\ref{fig:virtual_lab} and the interactive version
\url{https://lab.quantumflytrap.com/lab/quantum-teleportation}.

In \emph{Virtual Lab}, we can simulate quantum teleportation with a photon source (in which photon polarization can be set to an arbitrary state), with the nonlinear crystal generating a Bell pair. The Bell state measurement is implemented with a CNOT Gate, and two classical bits are sent to the remote location. 

For two qubits, there is only one type of entanglement. That is, each entangled state can be turned into another by applying only local operations and classical communication (LOCC). For three and more qubits, this is no longer the case\cite{migdal_entanglement_2013}. There are two distinct states that cannot be changed locally -- W and \textbf{Greenberger–Horne–Zeilinger (GHZ) states}\cite{greenberger_going_1989}, which can be used to test non-locality\cite{pan_experimental_2000}:
\begin{align}
    \ket{\text{W}} &= \frac{\ket{HHV} + \ket{HVH} + \ket{VHH}}{\sqrt{3}}\\
    \ket{\text{GHZ}} &= \frac{\ket{HHH} + \ket{VVV}}{\sqrt{2}}
\end{align}

\section{Usage}
\label{sec:usage}

University professors and educators use \emph{Virtual Lab} as a teaching tool. It has been used during quantum information courses at the \emph{University of Oxford}\cite{ekert_online_2022} and \emph{Stanford University}. We have learned from one professor, whose experimental lab was closed during the COVID pandemic, that \textit{“even when it becomes possible to teach my class in person again, your website could help me scale to a larger class size than is possible when I require equipment for hands-on demonstrations.”} Another educator told us that \textit{“we used Quantum Flytrap in labs over the past week and our students loved it! They really enjoyed the opportunity to play around with different optical elements. It also helped them visualize superposition, interference, wave-particle duality, and measurements.”}

\emph{Virtual Lab} has been used during several hackathons and other informal educational events and by individual users. As an easily embeddable tool, it has been used by external educational portals for quests and tutorials introducing quantum mechanics and quantum computing, including \emph{Qubit by Qubit}, \emph{QPlayLearn}\cite{foti_quantum_2021}, and \emph{Quantencomputer für Schüler\_innen}\cite{chaoui_quantecomputing_nodate}. We have approximately 70 unique users per day, growing to 700 during events. There are over 450 user-created experiments, which can be shared. We have created a Discord channel to gather valuable feedback, cultivate a discussion, and develop a new quantum community.
  
\emph{Virtual Lab} was shortlisted for \emph{D\&AD Pencils}, one of the most prominent awards in design\cite{noauthor_quantum_2021} and presented as a demo at a prestigious human-computer interaction conference \emph{ACM CHI 2022}\cite{jankiewicz_virtual_2022}. It is also listed in the \emph{Quantum Flagship} outreach resources section\\
\url{https://qt.eu/about-quantum-flagship/outreach/}.

\emph{Virtual Lab}, as well as its predecessors, \emph{Quantum Game with Photons} and \emph{Quantum Game with Photons 2}, have been mentioned in numerous peer-reviewed publications on games in quantum mechanics\cite{dorland_quantum_2019,ashktorab_thinking_2019,weisz_entanglion_2018}, teaching quantum literacy\cite{foti_quantum_2021}, general serious games\cite{parakh_approach_2019,parakh_novel_2020}, and preprints on computational complexity of puzzles\cite{costa_computational_2018} and the quantum workforce\cite{kaur_defining_2022}.

\section{Conclusion}
\label{sec:conclusion}

\emph{Virtual Lab by Quantum Flytrap} is a browser-based simulation of quantum mechanics of up to three photons. It provides novel ways of interacting with systems and setups in quantum optics. 

We designed new ways of visualizing vectors and matrices, the entanglement of arbitrary pure states, and measures of entanglement between particles. These tools are ready-to-use for arbitrary quantum systems, as we provide the open-source packages \emph{Quantum Tensors} and \emph{BraKetVue}. 

\emph{Virtual Lab} can be further developed by being extended to more photons, adding other degrees of freedom (e.g. Laguerre-Gaussian modes for angular optical momentum\cite{allen_optical_2003}), three spatial dimensions (for entertainment and interactive educational experiences with Virtual Reality), other carriers of quantum information (e.g. fermions such as electrons, electron states of ions and neutral atoms), and mixed states (represented by a density matrix, or measured via sampling).

Moreover, we can use the \emph{Quantum Flytrap} simulation engine to create no-code user interfaces to interact with quantum devices. The real-time simulation allows for testing and debugging quantum software in a way similar to a typical programming Integrated Development System (IDE). We believe that these end-user tools are necessary to make quantum computing accessible to a broad technical workforce of software engineers, data scientists, and analysts.

\subsection*{Acknowledgments}

We acknowledge the support of the \emph{Centre for Quantum Technologies}, the \emph{National University of Singapore} in 2019 -- the project started there as \emph{Quantum Game with Photons 2}, thanks to an invitation by Artur Ekert. We are particularly grateful to Evon Tan for all her administrative and organizational help and support.

PM acknowledges the support of \emph{eNgage – III/2014 grant by the Foundation for Polish Science}, for Q\emph{uantum Game with Photons} (2016), as well as fruitful discussions with its other creators Patryk Hes and Michał Krupiński. 
Developing \emph{BraKetVue} as a standalone, open-source matrix visualizer was supported by the \emph{Unitary Fund} microgrant.

A number of other people contributed to other than quantum aspects of \emph{Virtual Lab}, including a procedural soundtrack by Paweł Janicki and software engineering support, especially from Jakub Strebeyko. The project benefited immensely from multiple discussions with people from the explorable explanations community, especially Andy Hall, Eryk Kopczyński, Dorota Celińska, Laur Nita, and Nicky Case. We were encouraged by public and private feedback from Scott Aaronson, Paul G. Kwiat, Monika Schleier-Smith, and James Wootton.
We are grateful to Sarah Martin and Carrie Weidner for their valuable feedback and extensive editorial support.

And last but not least, we would like to thank all users, players, students, and educators. It is you who continuously motivated us to keep developing \emph{Virtual Lab}.


\bibliography{quantum-flytrap}   
\bibliographystyle{spiejour}   

\end{spacing}
\end{document}